\documentclass[a4paper]{jpconf}
\usepackage{graphicx}

\begin{document}
\title{Pseudoscalar-exchange contribution to $(g-2)_{\mu}$ from rational approximants}

\author{Pablo~Sanchez-Puertas$^{1}$ and Pere Masjuan$^{2}$}

\address{PRISMA Cluster of Excellence, Institut f\"ur Kernphysik, Johannes Gutenberg-Universit\"at, Mainz D-55099, Germany}

\ead{$^{1}$sanchezp@kph.uni-mainz.de,$^{2}$masjuan@kph.uni-mainz.de}

\begin{abstract}
We present our recent results on the hadronic light-by-light pseudoscalar-exchange contribution to the anomalous magnetic moment of the muon using rational approximants. Our work 
provides a generalization of Pad\'e approximants to the bivariate case to describe the most general doubly-virtual transition form factor, which is required in the 
calculation. This method provides a powerful tool to systematically implement the known experimental data on transition form factors in the space-like region and low energies 
in a model-independent way. Given the lack of experimental data on the doubly-virtual transition form factor, we make use of the pseduoscalar decays into lepton pairs. We find an 
interesting and puzzling situation which calls for new experimental measurements to clarify the present state. 
\end{abstract}

\section{Introduction}
%There has been a persistent discrepancy among the theoretical prediction and the experimental value for the anomalous magnetic moment of the muon, 
%$(g-2)_{\mu}$~\cite{Jegerlehner:2009ry}, which at the moment reads ($a_{\mu}\equiv (g-2)_{\mu}$)~\cite{Knecht:2014sea}
%%
%\begin{equation}
%a_{\mu}^{exp}-a_{\mu}^{th} = (116592091(63)-116591826(57))\times10^{-11}  =  265(85)\times10^{-11}
%\end{equation}
%%
%and may be an indication of new physics beyond the standard model. However, to reach to such conclusion, higher precision is required. For this reason there are two different 
%projected experiments at JPARC~\cite{Mibe:2010zz} and Fermilab~\cite{LeeRoberts:2011zz} aiming for a precision of $\mathcal{O}(10^{-10})$.\\

There has been a persistent discrepancy among the theoretical prediction and the experimental value for the anomalous magnetic moment of the muon, 
$a_{\mu}^{exp}=116592091(63)\times10^{-11}$ with $a_{\mu}\equiv (g-2)_{\mu}$, at the $3\sigma$ level~\cite{Jegerlehner:2009ry,Agashe:2014kda,Masjuan:2014rea} which
may call for physics beyond the standard model. However, to reach to such conclusion, higher precision is required. For this reason there are two different 
projected experiments at Fermilab~\cite{LeeRoberts:2011zz} and JPARC~\cite{Mibe:2010zz} aiming for a precision at around $10^{-10}$.\\

\begin{figure}
\centering
   \includegraphics[width=0.4\textwidth]{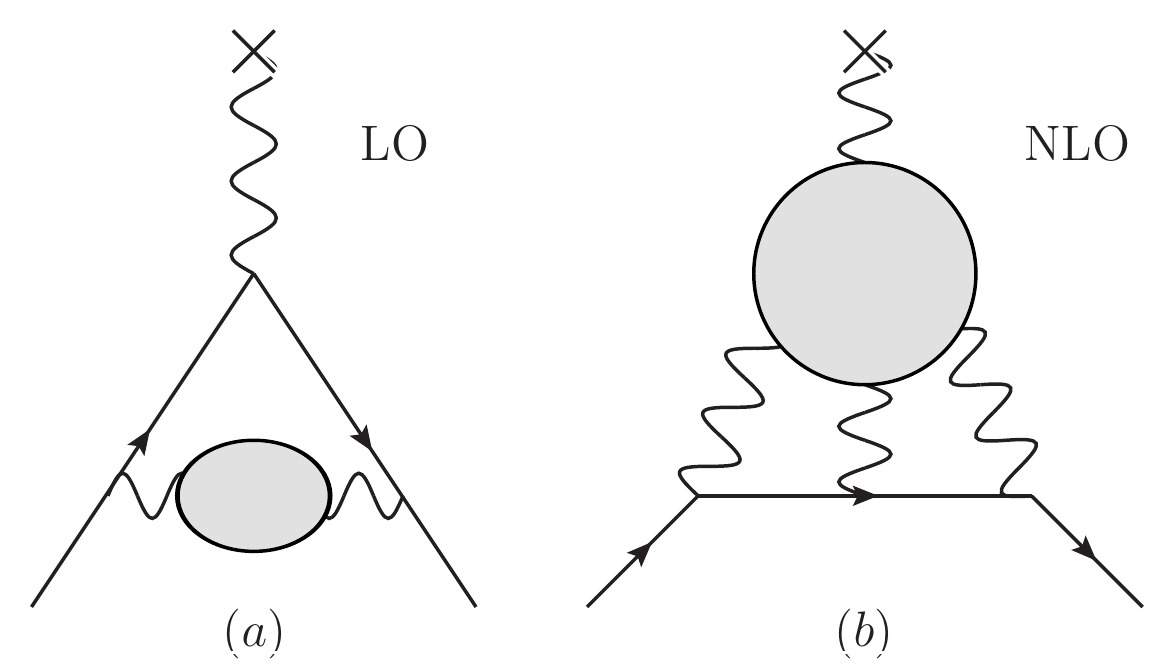} \hspace{1cm}
   \begin{minipage}[b]{0.4\textwidth}\caption{\label{fig:H2g} Hadronic contributions to $(g-2)_{\mu}$. Diagram $(a)$ represents the hadronic vacuum polarization, whereas 
   diagram $(b)$ stands for the hadronic light-by-light scattering contribution.} \end{minipage}
\end{figure}

Regretfully, the experimental effort will not be enough if it is not complemented by an equally precise theoretical prediction, which uncertainty is totally given by the hadronic 
contributions~\cite{Jegerlehner:2009ry}. However, these calculations are not an easy task as they involve quantum chromodynamics (QCD) at all energy 
scales, which represents a multi-scale problem and has required a close collaboration among theorists and experimentalists. 
A beautiful example is the leading hadronic contribution to $(g-2)_{\mu}$, the hadronic vacuum polarization (HVP), which is shown in Fig.~\ref{fig:H2g}a, where the blob stands 
for all-possible intermediate hadronic states. There, the optical theorem relates this quantity to experimentally-measurable cross sections 
$\sigma(e^+e^-\rightarrow\gamma^*\rightarrow \textrm{hadrons})$, and an ambitious experimental program promises to achieve the required precision for this quantity, which 
at present is of $43\times10^{-11}$~\cite{Knecht:2014sea}.
\\

Certainly, more complicated is the situation for the hadronic light-by-light (HLBL) contribution, which appears at NLO together with the NLO corrections to the HVP and is shown 
in Fig.~\ref{fig:H2g}b. This may become the dominant theoretical uncertainty, which at the moment is estimated as $40\times10^{-11}$~\cite{Knecht:2014sea}. In contrast to the 
HVP, the HLBL cannot be directly related to any measurable cross section, and again requires the knowledge of QCD at all scales, which is represented in Fig.~\ref{fig:H2g}b by the 
grey blob. 
Being a multi-scale problem, it is not possible to rely on a perturbative $\alpha_s$ expansion. However, it is still possible to find guidance on the chiral and large-$N_c$ expansions, which 
allowed in Ref.~\cite{deRafael:1993za} to decompose the HLBL in the different contributions depicted in Fig.~\ref{fig:counting}. In this scheme, the $\pi^0,\eta$ and $\eta'$ exchanges 
together with the (numerically subleading) $\pi^{\pm}$ and $K^{\pm}$ loops give the major contributions, see Tab.~2 in Ref.~\cite{Masjuan:2014rea}. 
This has triggered an ambitious program and great interest in the field of $\gamma\gamma$ physics.
Still, such calculations are complicated as there is not an obvious tool to implement all the available data and theoretical constraints. Often, one relies then on simplified models where 
the intrinsic errors are difficult to estimate. The current reference numbers, $116(39)\times10^{-11}$~\cite{Jegerlehner:2009ry} and $105(25)\times10^{-11}$~\cite{Prades:2009tw}, suffer 
indeed from this problem. 
In order to supply this shortcoming, different approaches have been proposed. As an example, lattice calculations~\cite{Blum:2014oka,Green:2015sra}, Dyson-Schwinger 
equations~\cite{Eichmann:2014ooa} and dispersive approaches~\cite{Colangelo:2014dfa,Colangelo:2014pva,Pauk:2014rfa} have been proposed. The last are based on the use of 
data at low energies, but lack the ability to implement the high-energy constraints. In our study, we propose a mathematical framework which allows to make full use of data and 
high-energy constraints and describe, in a model-independent fashion, the required quantities for calculating the $\pi^0,\eta$ and $\eta'$ contribution to the HLBL, which represents 
the dominant piece. For this, we extend the approach based on Pad\'e approximants described in Refs.~\cite{Masjuan:2012wy,Escribano:2013kba} to the double virtual case. For the 
moment, we have to face the situation that no double-virtual data is available. At this respect, we discuss in the last section what can be learnt from 
$P\rightarrow\overline{\ell}\ell$ decays, where $P=\pi^0,\eta,\eta'$.

%However, it is expected from chiral principles and large-$N_c$ arguments, that the $\pi^{\pm}$ and $K^{\pm}$ loops, together with the $\pi^0,\eta$ and $\eta'$ exchange 
%give the major contribution~\cite{deRafael:1993za}. This has triggered an ambitious program and great interest in the field of $\gamma\gamma$ physics {\textit{with renovated interest 
%after recent advances on dispersive approaches~\cite{Pauk:2014rfa,Colangelo:2014dfa,Colangelo:2014pva}}}. However, such calculations are still not easy as there is not an easy tool 
%to implement the data, which usually require some modelization procedure. In our study, we propose a model-independent approach which allows to make full use of data and describe, 
%in a model-independent fashion, the required quantities for calculating the $\pi^0,\eta$ and $\eta'$ contribution to the HLBL, which represents the dominant piece. Given the lack 
%of data, we discuss what can be indirectly learnt from $P\rightarrow\overline{\ell}\ell$ decays, where $P=\pi^0,\eta,\eta'$.

\section{Pseudoscalar-exchange hadronic light-by-light contribution to $(g-2)_{\mu}$}

\begin{figure}
\centering
   \includegraphics[width=0.42\textwidth]{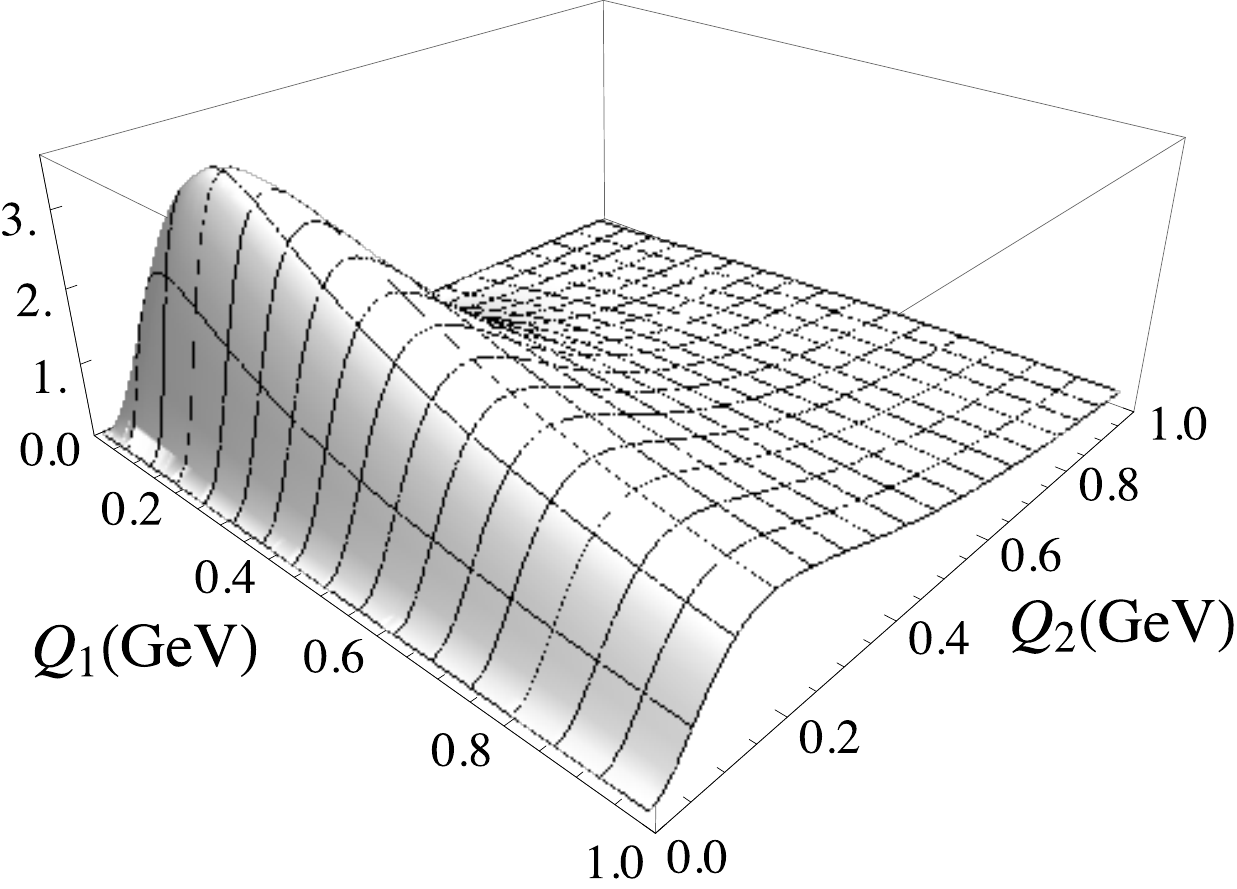}
   \includegraphics[width=0.42\textwidth]{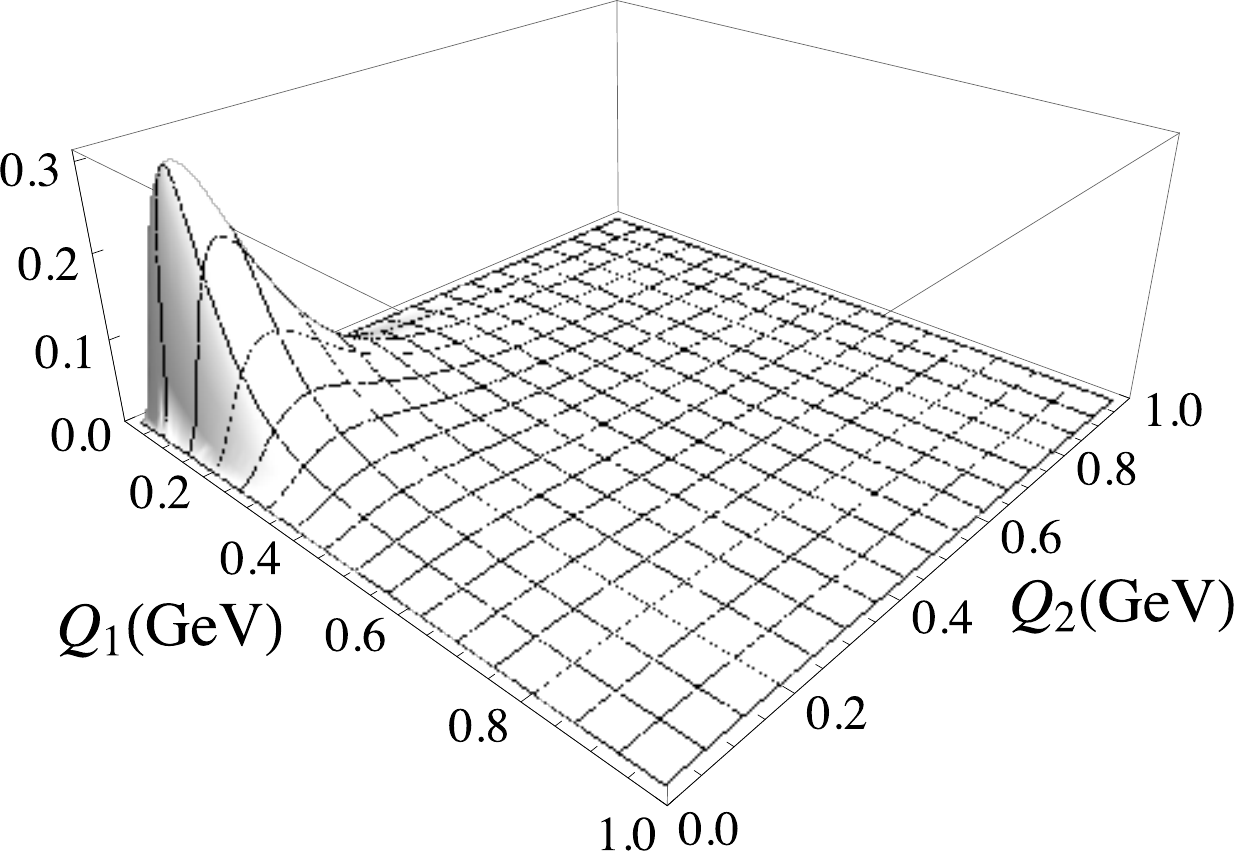}
   \caption{\label{fig:kernel} Integrals from Eq.~(\ref{eq:HLBL}) for a constant TFF and $t=0$ (the $(\alpha/\pi)^3$ factor is omitted). Left and right graphics represent the first and 
                 second term in the integral, respectively.}
\end{figure}

Given that the (light) pseudoscalar exchange represents the major contribution to the HLBL, such calculation becomes of central importance when aiming for precision.
%The HLBL contribution to $(g-2)_{\mu}$ is a rather tough quantity to calculate. Involving non-perturbative hadronic physics, we cannot use $\alpha_s$ as an expansion parameter. 
%However, we can still find guidance in the chiral and large-$N_c$ expansions~\cite{deRafael:1993za}. This argument made clear that the dominant contributions should be given by 
%the $\pi^{\pm}$ and $K^{\pm}$ loops and the $\pi^0,\eta$ and $\eta'$ exchange, see Fig.~\ref{fig:counting}. 
%
\begin{figure}[b]
\centering
  \includegraphics[width=0.8\textwidth]{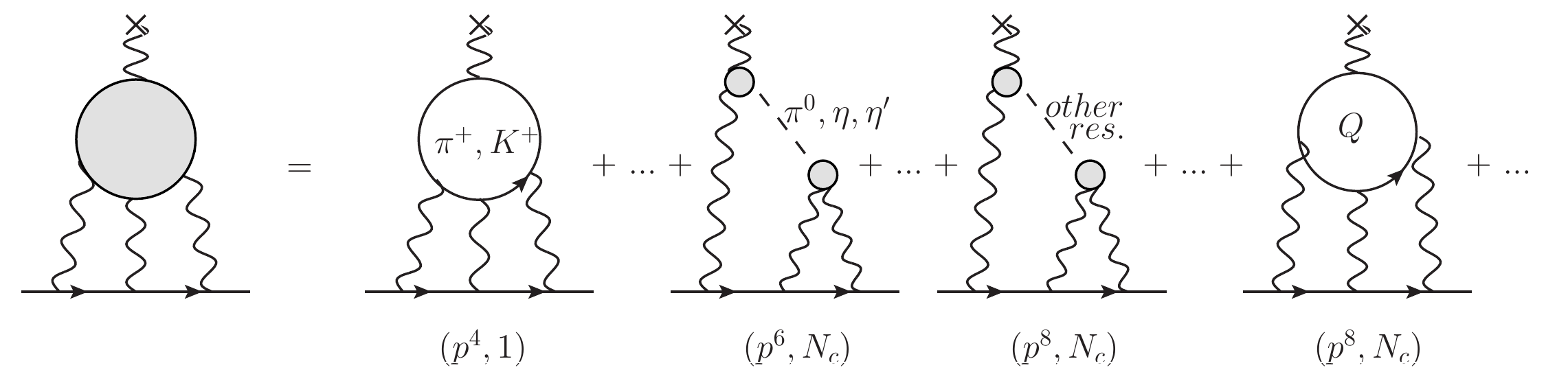}
  \caption{\label{fig:counting}The combined chiral and large-$N_c$ counting for HLBL proposed in Ref.~\cite{deRafael:1993za}.}
\end{figure}
%
%It turns out however that, phenomenologically, the pseudoscalar-exchange contribution 
%dominates~\cite{Jegerlehner:2009ry}. Therefore, we need to provide a precise calculation for this quantity, which requires an accurate description of $P\gamma^*\gamma^*$ interactions, 
%which are described in terms of the pseudoscalar transition form factor (TFF) $F_{P\gamma^*\gamma^*}(q_1^2,q_2^2)$ as
This requires an accurate description of $P\gamma^*\gamma^*$ interactions, which are represented as the grey blobs in Fig.~\ref{fig:counting}, and are described in terms of the 
pseudoscalar transition form factor (TFF) $F_{P\gamma^*\gamma^*}(q_1^2,q_2^2)$ as
\begin{equation}
i\mathcal{M} = i\epsilon_{\mu\nu\rho\sigma}\epsilon_1^{\mu}q_1^{\nu}\epsilon_2^{\rho}q_2^{\sigma}F_{P\gamma^*\gamma^*}(q_1^2,q_2^2),
\end{equation}
meaning that the $\pi^0$ (as well as the $\eta$ and $\eta'$) TFF $F_{\pi^0\gamma^*\gamma^*}(q_1^2,q_2^2)$ will play an essential role in determining this quantity. Actually, the 
contribution to $a_{\mu}$ can be expressed in terms of this as~\cite{Jegerlehner:2009ry}
\begin{equation}
\label{eq:HLBL}
a_{\mu}^{HLBL;P} =  \frac{-2\pi}{3}\left( \frac{\alpha}{\pi} \right)^{3} \int_0^{\infty}dQ_1dQ_2 \int_{-1}^{+1}dt \sqrt{1-t^2} Q_1^3Q_2^3  
                            \left[  \frac{F_1 I_1(Q_1,Q_2,t)}{Q_2^2+m_{P}^2}  +    \frac{F_2 I_2(Q_1,Q_2,t)}{Q_3^2+m_{P}^2}  \right],
\end{equation}
where the $I_i(Q_1,Q_2,t)$ functions may be found in Ref.~\cite{Jegerlehner:2009ry}, $Q_3^2 = Q_1^2+Q_2^2 + Q_1Q_2t$, and the TFF appears through the quantities
\begin{equation}
\label{eq:HLBLFF}
F_1 = F_{P\gamma^*\gamma^*}(-Q_1^2,-Q_3^2)F_{P\gamma^*\gamma}(-Q_2^2,0), \quad F_2 = F_{P\gamma^*\gamma^*}(-Q_1^2,-Q_2^2)F_{P\gamma^*\gamma}(-Q_3^2,0), 
\end{equation}
with $Q_i^2$ a space-like variable and $t$ an angular one. The behavior for this integral is shown in Fig.~\ref{fig:kernel} for $t=0$, though very similar shape is obtained for other values. 
A relevant observation is that the previous integral is peaked at very low space-like energies. Aiming for very small errors, smaller than $10\%$, it is extremely important 
to provide a precise description at 
these scales around $(0-1)$~GeV. In addition, specially for the first term in the integral in Eq.~(\ref{eq:HLBL}) which is UV-divergent for a point-like TFF, it is important to provide the adequate 
high-energy behavior dictated by perturbative QCD ---which is the drawback in dispersion relation approaches. %--- whereas the second subleading term, UV-finite, is rather insensitive to this. 
This makes a clear statement that, for reaching a precise determination for this quantity, we need to provide a precise description of the pseudoscalars TFFs at very low-energies 
together with the appropriate high-energy behavior, which are the main concerns of our approach discussed below.

\section{A rational description for $F_{P\gamma^*\gamma^*}(-Q_1^2,-Q_2^2)$}

Achieving a TFF description from first principles is extremely challenging, if not impossible, at the moment. On the one hand, at the very low energies involved in the process, one could 
think about resorting to a chiral perturbation theory ($\chi$PT) approach. However, any description beyond a constant TFF requires many unknown low-energy constants and violates 
unitarity at high energies. Moreover, the convergency of this expansion breaks down much before the first resonance scale, still relevant in our integral. For details on a 
$\chi$PT-based calcualtion for the HLBL, see Ref.~\cite{RamseyMusolf:2002cy}. 
On the other hand, for very large energies, we can rely on perturbative QCD. In this framework, the TFF is obtained in terms of a convolution of a hard-scattering amplitude with 
the pseudoscalar distribution amplitude~\cite{Lepage:1980fj}. The latter object, which is of non-perturbative nature, encodes the $\pi^0$ structure and must be modeled, inducing 
potential large uncertainties. Consequently, the only reliable limits at disposal are~\cite{Lepage:1980fj,Melnikov:2003xd,Knecht:2001xc}
\begin{equation}
\label{eq:tffhe}
   \lim_{Q^2\to\infty} Q^2F_{P\gamma^*\gamma}(-Q^2,0) = 2F_{\pi}, \quad    \lim_{Q^2\to\infty} Q^2F_{P\gamma^*\gamma^*}(-Q^2,-Q^2) = (2/3)F_{\pi},
\end{equation}
where $F_{\pi}=92$~MeV is the pion decay constant~\cite{Agashe:2014kda}. Therefore, it is necessary to find an alternative approach able to link both, the low and the high 
energies as well as controlling the theoretical uncertainties. 
For this, some models, such as hidden local symmetry~\cite{Benayoun:2015gxa}, resonant chiral theory~\cite{Roig:2014uja}, or large-$N_c$ based generalizing vector meson dominance 
ideas~\cite{Knecht:2001qf,Husek:2015wta} have been proposed. Their qualitative agreement may be understood from the large-$N_c$ limit of QCD, where Green's functions are expressed 
in terms of an infinite number of resonances exchange~\cite{'tHooft:1973jz,Witten:1979kh}. They lack however the estimation of the theoretical error which is associted to their 
model or simplifying assumptions, and the deeper question remains on whether it would be possible in these models to reproduce the TFF to arbitrary precision. It was proposed 
however in Ref.~\cite{Masjuan:2007ay}, that the agreement of this approaches 
can be understood as well from the framework of Pad\'e theory. The advantage of the latter is that, for certain kind of functions, it allows to go beyond the large-$N_c$ limit, 
and approximate the real world QCD function. Actually, these properties have already been exploited for dealing with the HVP~\cite{Masjuan:2009wy,Aubin:2012me} and it has been 
shown that they allow to estimate the theoretical uncertainties. We propose to use this framework for reproducing the TFF arising in the pseudoscalar exchange contribution to HLBL as 
well~\cite{Masjuan:2012wy,Escribano:2013kba}.\\

Given an analytic function with known series expansion, say, $F_{P\gamma^*\gamma}(-Q^2,0) = F_{P\gamma\gamma}(0,0)(1-b_PQ^2+c_PQ^4 + ...)$, Pad\'e approximants (PA) are 
rational functions of two polynomials, $R_N(Q^2)$ and $Q_M(Q^2)$ of degree $N$ and $M$, respectively, constructed such as to match the original series expansion up to order $\mathcal{O}((Q^2)^{N+M+1})$ terms~\cite{baker}, 
\begin{equation}
\label{eq:pa}
  P^N_M(Q^2) = \frac{R_N(Q^2)}{Q_M(Q^2)} = F_{P\gamma\gamma}(0,0)[1-b_PQ^2+c_PQ^4 + ... + \mathcal{O}((Q^2)^{N+M+1})].
\end{equation}
Pad\'e approximants can be proven to approximate not only meromorphic functions ---which represents the large-$N_c$ limit of QCD--- but Stieltjes functions as well~\cite{baker}. 
These kind of functions may have threshold discontinuities and often represents cases of physical interest~\cite{Masjuan:2009wy}. Pad\'e theory guarantees then the convergence 
of some kind of sequences $P^{N}_{N+M}(Q^2)$ in the whole complex-plane except for the discontinuity itself, where the original function is ill-defined. These theorems 
provide then a mathematical corpus of the model-indepedendency of our approach as the known analytic structure of TFF seems to fall in this kind of 
functions~\cite{Escribano:2015nra}. 
In this respect, we are going a step further with respect to previous approaches, guaranteeing the ability to reproduce, to arbitrary precision, the TFF. Note that this holds 
for some previous approaches in the strict large-$N_c$ limit of QCD alone~\cite{Masjuan:2007ay}.
Moreover, the systematic construction, Eq.~(\ref{eq:pa}) allows to estimate the theoretical 
uncertainty~\cite{Masjuan:2012wy,Masjuan:2008fv,Masjuan:2009wy,Escribano:2013kba,Escribano:2015nra}. \\

Still, it is necessary to provide the low-energy expansion, Eq.~(\ref{eq:pa}), before we can apply the method. In this respect, it was shown in Ref.~\cite{Masjuan:2008fv}, that 
PAs can be used as well for this purpose in order to safely extract the low-energy parameters from experimental data for the space-like and low-energy time-like TFF. 
In Refs.~\cite{Masjuan:2009wy,Escribano:2013kba,Escribano:2015nra}, we were able to extract in this way some of the leading parameters for the $\pi^0,\eta$ and $\eta'$, 
allowing us to reconstruct the first approximants and obtain valuable information, as the $\eta$ and $\eta'$ mixing parameters. This allows to safely reconstruct the 
TFFs at that low-energies where often no data is available.\\

So far, this provides a description for the single virtual TFF $F_{P\gamma^*\gamma}(Q^2,0)$. However, from Eqs.~(\ref{eq:HLBL})~and~(\ref{eq:HLBLFF}), we find that it is the 
most general doubly-virtual TFF $F_{P\gamma^*\gamma^*}(Q^2,Q^2)$ that we require  in our calculation. To describe this, we need to extend the PAs to the bivariate case for 
symmetric functions, which are known as Canterbury approximants (CA), see Ref.~\cite{Masjuan:2015lca} and references therein. 
They are constructed, similarly to PA, from the original series expansion and the simplest approximant reads~\cite{Masjuan:2015lca}
\begin{equation}
\label{eq:ca}
  C^0_1(Q_1^2,Q_2^2) = \frac{F_{P\gamma\gamma}(0,0)}{1+b_P(Q_1^2+Q_2^2) + (2b_P^2-a_{1,1})Q_1^2Q_2^2},
\end{equation}
with $b_P$ the TFF slope and $a_{1,1}$ given by the doubly-virtual expansion $F_{P\gamma^*\gamma^*}(-Q_1^2,-Q_2^2) = F_{P\gamma\gamma}(0,0)(1-b_P(Q_1^2+Q_2^2) + a_{1,1}Q_1^2Q_2^2 
+ ...)$. As already discussed in the introduction, there is at present not available experimental data for the 
doubly-virtual TFF, which would allow to extract the low-energy parameters in a similar manner as for the single-virtual TFF. This means that we cannot go to large approximants. 
Then, we stick for the moment to the simplest element Eq.~(\ref{eq:ca}) and judge, based on theoretical constraints, a reasonable range in which the real $a_{1,1}$ parameter would 
lie be contained. On the one side, at low energies, $\chi$PT offers indications that the TFF should factorize~\cite{Bijnens:2012hf}, i.e., 
$F_{P\gamma^*\gamma^*}(-Q_1^2,-Q_2^2)\sim F_{P\gamma^*\gamma}(-Q_1^2,0) \times F_{P\gamma^*\gamma}(-Q_2^2,0)$, 
implying then $a_{1,1} = b_P^2$. On the other side, at large energies, the OPE condition, 
Eq.~(\ref{eq:tffhe}), suggest that $a_{1,1}=2b_P^2$ so that the high-energy behavior is fulfilled. Guided by this hints, we choose an even more conservative range 
$a_{1,1} \in \{0,2b_P^2\}$~\cite{Masjuan:2015lca}, which we believe is generous enough to consider possible departure from factorization at higher energies\footnote{ The 
factorization assumption at low energies used in Ref.~\cite{Masjuan:2015lca} for estimating our $a_{1,1}$ range was confirmed in recent dispersive analysis for the $\eta$ case~\cite{Xiao:2015uva}}. 
Of course, ultimately, experimental data will have the last word on this. Let us remark though, that either of the above-mentioned choices have small impact at low energies 
---see discussion from the authors in the proceedings of Chiral Dynamics 2015.

\section{Results for $a^{HLBL;\pi^0\eta,\eta'}_{\mu}$}
With the doubly-virtual behavior incorporated, we can come back to the HLBL contribution.
Using our chosen band for the double virtual parameter $a_{1,1}$, which we stress, not only considers the correct low-energy behavior, but account as well for the high-energy 
one (i.e., it reproduces for $a_{1,1}=2b_P^2$ the  power-like behavior in Eq.~(\ref{eq:tffhe})), we obtain the preliminary results for their contribution to HLBL, 
$a^{HLBL;\pi^0\eta,\eta'}_{\mu}$, in Table~\ref{tab:g2res}.
\begin{table}
\caption{  \label{tab:g2res} $a^{HLBL;\pi^0\eta,\eta'}_{\mu}$ preliminary results for different values in the chosen $a_{1,1}$ range.}
\begin{center}
  \begin{tabular}[t]{lllll} 
  \br
       Units of $10^{-10}$      & $\pi^0$ & $\eta$ & $\eta'$  &  Total \\ 
  \mr
   {\small{$a_{1,1} = 2b_{P}^2 \ \ [OPE]$}} \hspace{0.2cm}   & $6.64(33)$ & $1.69(6)$ & $1.61(21)$  &  $9.94(40)_{stat}(50)_{sys}$ \\
   {\small{$a_{1,1} = b_{P}^2$ \ \ [Fact]}}    & $5.53(27)$ & $1.30(5)$ & $1.21(12)$  &  $8.04(30)_{stat}(40)_{sys}$ \\
   {\small{$a_{1,1} = 0$}}   & $5.10(23)$ & $1.16(7)$ & $1.07(15)$  &  $7.33(28)_{stat}(37)_{sys}$ \\ 
  \br
  \end{tabular}
  \end{center}
\end{table}
There, we quote the statistical error from the data-based procedure together with our estimated $5\%$ systematic error~\cite{Escribano:2013kba}. For the first 
time, a fully data-driven with high-energy constraints implementation and systematic errors has been provided.
We remark that OPE behavior was not considered for the $\eta$ and $\eta'$ cases 
in the reference numbers from Refs.~\cite{Jegerlehner:2009ry,Prades:2009tw} which used a factorization approach based on resonance ideas which are not required in our framework. 
We find that the error on this assumption may not be negligible, i.e. at the order of the projected experimental precision $\sim 16\times10^{-11}$. 
The overall uncertainty is dominated by the $\pi^0$ contribution, whereas the $\eta$ result has been greatly improved as compared to 
Ref.~\cite{Escribano:2013kba}. This has been possible thanks to the new low-energy parameters extraction in Ref.~\cite{Escribano:2015nra} including, for the first time, not 
only the space-like, but the time-like data from the precise Dalitz decay 
measurements from A2~\cite{Aguar-Bartolome:2013vpw} collaboration at Mainz. This illustrates the potentiality of the method to benefit from very different   
experimental regimes. In principle, new precise data from the $\pi^0$ TFF would allow for a similar improvement. In this respect, we are currently working on the impact of new 
data on these errors~\cite{gm2}. Accounting for our quoted range we obtain that~\cite{gm2} 
\begin{equation}
 a^{HLBL;\pi^0\eta,\eta'}_{\mu} = (73.3 \div 99.4)(\pm6.4)\times10^{-11},
\end{equation}
where the lower(upper) values come from $a_{1,1}=0(2b_P^2)$ choice. We stress that, at this point, it is not possible to choose either factorization or OPE as this would 
be in contradiction either with the high- or low-energy behavior. 
This is a peculiarity of the lowest approximation and calls for the use of the $C^1_2(Q_1^2,Q_2^2)$ approximant, which already has the ability to 
implement both behaviors~\cite{gm2}. 
We remark that we do not consider the so called off-shell effects~\cite{Jegerlehner:2009ry,Melnikov:2003xd} in our calculation. See Ref.~\cite{Masjuan:2012qn} in this respect.
So far, this represents our best estimate, which 
global uncertainty is clearly dominated by the chosen $a_{P;1,1}$ window, arising from our 
ignorance about the doubly-virtual TFF. This is greater than our uncertainty estimation, $\pm6.3\times10^{-11}$, and twice as big as the projected experimental $(g-2)_{\mu}$ uncertainties $\sim16\times10^{-11}$. 
This clearly illustrates the necessity of measuring the doubly-virtual TFF if we aim to achieve a similar precision as projected experiments. This
should answer the question on how fast the TFF reaches the (doubly-virtual) high-energy behavior and would allow for the $C^1_2(Q_1^2,Q_2^2)$ determination. Given the lack 
of knowledge both from the experimental side, where no data is available, and the theoretical side, which offer no answer yet, we consider in the next section the 
$P\rightarrow\overline{\ell}\ell$ decays, which as we argue, offer an indirect probe to this question.

\section{Pseudoscalar decays into a lepton pair and $(g-2)_{\mu}$ implications}

\begin{figure}
\centering
\includegraphics[width=0.5\textwidth]{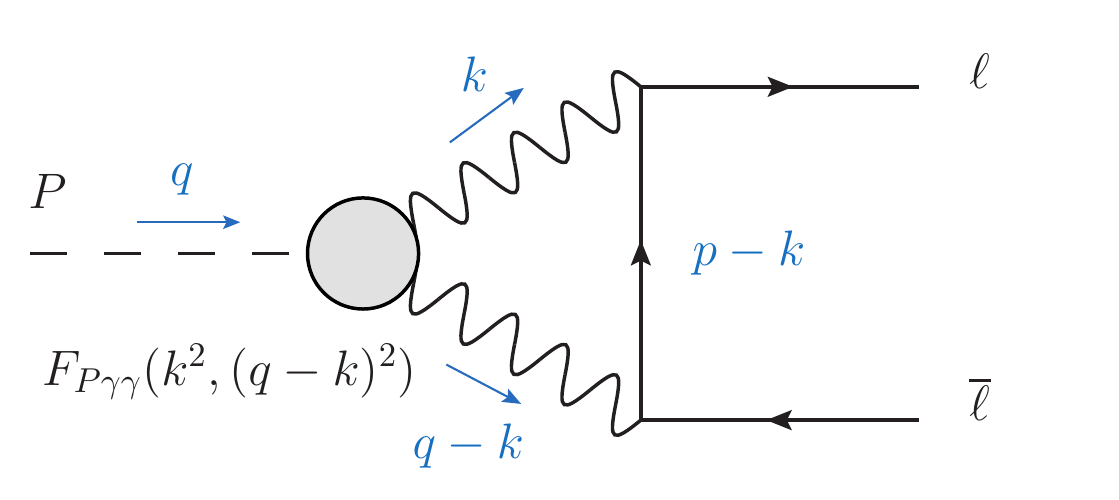} \hspace{1cm}
   \begin{minipage}[b]{0.4\textwidth}\caption{\label{fig:pll} The leading order QED contribution to the $P\rightarrow\overline{\ell}\ell$ process where $P=\pi^0,\eta,\eta'$. 
   $F_{P\gamma^*\gamma^*}(k^2,(q-k)^2)$ stands for the transition form factor.} \end{minipage}
\end{figure}

Whereas experimentally it is very difficult to probe the doubly-virtual TFF, limited by the low cross sections together with current achieved luminosities, there is an alternative indirect 
path to probe this. This possibility is brought, in analogy to new physics in $(g-2)_{\mu}$, through loop mediated processes which, being sensible to the TFF at the whole energy scale,  
provide us the opportunity to test high-energy effects in low-energy phenomena. In this case such possibility is brought by the pseudoscalar decays into lepton pairs,  
$P\rightarrow\overline{\ell}\ell$, for which $P=\pi^0,\eta,\eta'$. This process appears at leading order through the diagram shown in Fig.~\ref{fig:pll}.
As such, the TFF must be integrated at all energies offering the desired indirect probe.
The branching ratio (BR) for this decay may be expressed in terms of the two photon decay width as 
\begin{equation}
\frac{\textrm{BR}(P\rightarrow\overline{\ell}\ell)}{\textrm{BR}(P\rightarrow\gamma\gamma)} = 2 \left( \frac{\alpha m_{\ell}}{\pi m_P}  \right)^2 \beta_{\ell} \left\vert\mathcal{A}(m_P^2)\right\vert^2,
\end{equation}
where $\mathcal{A}(q^2)$ represents the loop amplitude (see Ref.~\cite{Masjuan:2015lca} and references therein for details)
\begin{equation}
\label{eq:loop}
\mathcal{A}(q^2) =  \frac{2i}{\pi^2q^2} \int d^4k \frac{ q^2k^2 - (q\cdot k)^2}{ k^2(q-k)^2((p-k)^2-m_{\ell}^2)}\tilde{F}_{P\gamma^*\gamma^*}(k^2,(q-k)^2), 
\end{equation}
which is unknown as far as the normalized TFF $(\tilde{F}_{P\gamma\gamma}(0,0)=1)$ is unspecified. The role of the doubly virtual TFF is actually rather important as the 
given integral, similarly to the HLBL case, is UV-divergent. Remarkably, for the $\pi^0$ case, it is possible to go further without a single clue on the TFF. Being the 
lightest hadron, it is not possible to find any additional intermediate hadronic 
state which may be on-shell, and contribute therefore to the imaginary part. Consequently, its imaginary part is solely given by the intermediate $\gamma\gamma$ state, which gives
\begin{equation}
   \textrm{Im}(\mathcal{A}(m_{\pi^0}^2)) =  \frac{\pi}{2\beta_{\ell}}\ln\left( \frac{1-\beta_{\ell}}{1+\beta_{\ell}} \right) \quad (\beta_{\ell}=\sqrt{1-4m_{\ell}^2/m_{\pi^0}^2}),
\end{equation}
inducing the so called unitary bound~\cite{Drell:1959} for the BR, $|\mathcal{A}(m_P^2)| \geq Im(\mathcal{A}(m_{\pi^0}^2))$, which for the $\pi^0$ gives 
$\textrm{BR}(\pi^0\rightarrow e^+e^-)=4.7\times10^{-8}$. 
This provides a model-independent lower bound, which violation would certainly imply physics beyond the standard model. \\

Experimentally it is found that, after removing the radiative corrections~\cite{Abouzaid:2006kk,Dorokhov:2007bd},
\begin{equation}
\textrm{BR}^{KTeV}(\pi^0\rightarrow e^+e^-) = 7.48(38)\times10^{-8} > \textrm{BR}^{th}(\pi^0\rightarrow e^+e^-) = 6.23(09)\times10^{-8}, 
\end{equation}
which is certainly (quite) above the unitary bound. Still, latest results on RC~\cite{Vasko:2011pi,Husek:2014tna} suggest 
$\textrm{BR}^{KTeV}(\pi^0\rightarrow e^+e^-) = 6.87(38)\times10^{-8}$. 
Nevertheless, they cannot yet justify such deviation, which would still imply some kind of new physics and has attracted much attention~\cite{Kahn:2007ru,Chang:2008np}. \\

Often, the unitary bound has been extended to the $\eta$ and $\eta'$ cases in order to provide an estimate for 
experimental searches. A word of caution comes here. While for the $\pi^0$ this was a firm result, this is not the case for the heavier 
$\eta$ and $\eta'$ mesons, which masses allow to include hadronic intermediate states such as $\pi^+\pi^-\gamma$ or even $\omega\gamma$ for the heavy $\eta'$.  
We find small corrections for the $\eta$ to the unitary bound, but large corrections to the $\eta'$ as the imaginary part is reduced by $20\%$ 
(see Ref.~\cite{etaetapdecays} and our proceedings in Chiral Dynamics 2015).\\

To go further and provide an estimation for these processes, we need to input some description for the TFF. Then, checking with the experimental values, we can find whether 
our description, assuming that new physics do not play any role here, corresponds to the experimental world. If this turns out not to be the case, it is possible that our 
 understanding of the doubly-virtual TFF is not as good as we believed. Once more, the crucial observation in this process is the dominance of the low-energy space-like 
$Q_1^2\simeq Q_2^2$ region which fully dominates the integral, see Ref.~\cite{Masjuan:2015lca}. This process is perfectly suited then for our CA description. Taking our 
quoted $a_{P;1,1}$ range, we obtain the preliminary results in Table~\ref{tab:pll}.
\begin{table}
\caption{\label{tab:pll}Our preliminary results for BR$(P\rightarrow\overline{\ell}\ell)$ for $a_{1,1}=(b_P^2,2b_P^2)$ range.}
\begin{center}
\lineup
  \begin{tabular}{llll} 
  \br
    Process & BR$^{th}$ & BR$^{exp}$ & Ref. \\ 
  \mr
    $\pi^0\rightarrow e^+e^-$ & $(6.20\div6.35)(5)\times10^{-8}$ & $7.48(38)\times10^{-8}$ & \cite{Abouzaid:2006kk}\\ 
    $\eta\rightarrow e^+e^-$ & $(5.31\div5.44)(^{+4}_{-5})\times10^{-9}$ & $\leq2.3\times10^{-6}$ & \cite{Agakishiev:2013fwl} \\
    $\eta\rightarrow \mu^+\mu^-$ & $(4.72\div4.52)(^{+4}_{-8})\times10^{-6}$ & $5.8(8)\times10^{-6}$ & \cite{Abegg:1994wx} \\ 
    $\eta'\rightarrow e^+e^-$ & $(1.82\div1.86)(19)\times10^{-10}$ & $\leq5.6\times10^{-9}$ & \cite{Achasov:2015mek,Akhmetshin:2014hxv} \\
    $\eta'\rightarrow \mu^+\mu^-$ & $(1.36\div1.49)(33)\times10^{-7}$ & ---&--- \\ 
  \br
  \end{tabular}
\end{center}
\end{table}
We provide the combination of both, statistical and systematic errors. For details on these, see Refs.~\cite{Masjuan:2015lca,etaetapdecays}.
We find that, for the $\pi^0$, previous results have probably underestimated their theoretical errors when modeling the doubly-virtual TFF~\cite{Dorokhov:2007bd}. 
%Still, our result lies $n(m)\sigma$ below the experimental measurement without(with) the latest RC incorporated. 
Actually, this feature applies for the $\eta$ and $\eta'$ cases 
as well~\cite{Dorokhov:2007bd}, where not only the double virtuality, but the single-TFF 
have been crudely modeled~\cite{Dorokhov:2009xs}. Moreover, we find for the $\eta$ and $\eta'$ remarkable deviations with respect to previous 
results~\cite{Dorokhov:2007bd,Knecht:1999gb} arising from 
the approximations adopted in previous calculations~\cite{Masjuan:2015lca,Dorokhov:2009xs}, which cannot be neglected for the $\eta$ and $\eta'$ decays. Finally, it is worth 
reminding that integral~(\ref{eq:loop}) is sensible as well to the time-like region up to the $m_P$ mass. This feature is of special importance for the $\eta'$. It can be shown that our method 
is able to effectively reproduce threshold effects there and we are safe to perform such calculation~\cite{etaetapdecays}, which is an important and distinctive feature 
in our approach not implemented so far.\\

We find from our results that both the $\pi^0\rightarrow e^+e^-$ and the $\eta\rightarrow\mu^+\mu^-$ decays show deviations from their experimental values, 
at $2.9\sigma$ and $1.3\sigma$, respectively. Remarkably, the latest bounds on $\eta'\rightarrow e^+e^-$~\cite{Achasov:2015mek,Akhmetshin:2014hxv} are reaching our 
predictions ---see the talks at this conference. Therefore, we encourage our experimental colleagues in Novosibirsk to further pursue this decay. Note that including 
radiative corrections alleviate the first result, leading to $1.5\sigma$. Yet its statistical significance, that represents a potential large deviation and, for the 
moment, the only clue about the doubly-virtual TFF effects. Given that they are dominated by the low-energy region, such effect would be encoded in the lowest-energy double virtual 
parameter $a_{1,1}$. Therefore, we can set it free and match to the experimental values. For the $\pi^0$, it requires a large negative $a_{1,1}$ value, greatly damping the 
TFF at low energies. In contrast, for the $\eta$ case this requires an extremely softly-falling TFF, which would require the use of the $C^1_2(Q_1^2,Q_2^2)$ approximant. 
Considering both results represents then a puzzling situation. Using this input to calculate the  $a^{HLBL;\pi^0}_{\mu}$ contribution, we obtain without(with) the latest RC results
\begin{equation}
 a^{HLBL;\pi^0}_{\mu} = 1.3(2.8)\times10^{-10}.
\end{equation}
This represents a large shift with respect to our previous prediction in Table~\ref{tab:g2res}, much beyond our quoted error. We conclude therefore that  such situation calls 
for an urgent revision. First, some experimental data would be required in order to improve our knowledge of the TFF. Second, a new precise measurement, with the latest 
results on radiative corrections accounted for, should be pursued. Only this would clarify the current situation and conclude whether we have the situation under control or, perhaps, 
new physics are playing some role in these decays.

\section{Conclusions and Outlook}

We have presented a model-independent approach based on rational approximants, namely, Canterbury approximants, to describe the pseudoscalar TFFs. This method provides 
both an useful tool to extract valuable information about TFF from experimental data as well as to reconstruct the TFF itself for later calculations. We have used this model to calculate 
then the $\pi^0,\eta$ and $\eta'$ HLBL 
contribution to $(g-2)_{\mu}$. We have shown that the current limitation comes from our uncertainty on the double-virtual TFF, for which no data is available yet. To supply this situation, 
we have made use of $P\rightarrow\overline{\ell}\ell$ decays. We have found that current experimental values present a puzzling situation which challenge our TFF description. 
Nevertheless, experimental work is required before any claim, which is specially urgent given the projected accuracy of future $(g-2)_{\mu}$ experiments. 
Meanwhile, to improve our description, we are studying the next $C^1_2$ approximant, which would allow to implement both, the low- and high-energy constraints at once and, we 
hope, will provide a better understading of the situation.

\section{Acknowledgments}

We thank Marc Vanderhaeghen for encouragements and discussions. Besides, P.~Sanchez-Puertas would like to thank the organizers 
for their hospitality and the opportunity to participate in this workshop  and enjoy from the inspiring environment there. Work supported by the Deutsche Forschungsgemeinschaft DFG 
through the Collaborative Research Center ``The Low-Energy Frontier of the Standard Model'' (SFB 1044), eprint: MITP/15-090.

\section*{References}

\providecommand{\newblock}{}


\begin{thebibliography}{10}
\expandafter\ifx\csname url\endcsname\relax
  \def\url#1{{\tt #1}}\fi
\expandafter\ifx\csname urlprefix\endcsname\relax\def\urlprefix{URL }\fi
\providecommand{\eprint}[2][]{\url{#2}}

\bibitem{Jegerlehner:2009ry}
Jegerlehner F and Nyffeler A 2009 {\em Phys. Rept.\/} {\bf 477} 1--110
  (\textit{Preprint} \eprint{0902.3360})

\bibitem{Agashe:2014kda}
Olive K~A {\em et~al.\/} (Particle Data Group) 2014 {\em Chin. Phys.\/} {\bf
  C38} 090001

\bibitem{Masjuan:2014rea}
Masjuan P 2015 {\em Nucl. Part. Phys. Proc.\/} {\bf 260} 111--115
  (\textit{Preprint} \eprint{1411.6397})

\bibitem{LeeRoberts:2011zz}
Lee~Roberts B (Fermilab P989) 2011 {\em Nucl. Phys. Proc. Suppl.\/} {\bf 218}
  237--241

\bibitem{Mibe:2010zz}
Mibe T (J-PARC g-2) 2010 {\em Chin. Phys.\/} {\bf C34} 745--748

\bibitem{Knecht:2014sea}
Knecht M 2015 {\em Nucl. Part. Phys. Proc.\/} {\bf 258-259} 235--240
  (\textit{Preprint} \eprint{1412.1228})

\bibitem{deRafael:1993za}
de~Rafael E 1994 {\em Phys. Lett.\/} {\bf B322} 239--246 (\textit{Preprint}
  \eprint{hep-ph/9311316})

\bibitem{Prades:2009tw}
Prades J, de~Rafael E and Vainshtein A 2009 {\em Adv. Ser. Direct. High Energy
  Phys.\/} {\bf 20} 303--317 (\textit{Preprint} \eprint{0901.0306})

\bibitem{Blum:2014oka}
Blum T, Chowdhury S, Hayakawa M and Izubuchi T 2015 {\em Phys. Rev. Lett.\/}
  {\bf 114} 012001 (\textit{Preprint} \eprint{1407.2923})

\bibitem{Green:2015sra}
Green J, Gryniuk O, von Hippel G, Meyer H~B and Pascalutsa V 2015
  (\textit{Preprint} \eprint{1507.01577})

\bibitem{Eichmann:2014ooa}
Eichmann G, Fischer C~S, Heupel W and Williams R 2014  {\em {11th
  Conference on Quark Confinement and the Hadron Spectrum (Confinement XI) St.
  Petersburg, Russia, September 8-12, 2014}\/} (\textit{Preprint}
  \eprint{1411.7876})
  \urlprefix\url{https://inspirehep.net/record/1331422/files/arXiv:1411.7876.pdf}

\bibitem{Colangelo:2014dfa}
Colangelo G, Hoferichter M, Procura M and Stoffer P 2014 {\em JHEP\/} {\bf 09}
  091 (\textit{Preprint} \eprint{1402.7081})

\bibitem{Colangelo:2014pva}
Colangelo G, Hoferichter M, Kubis B, Procura M and Stoffer P 2014 {\em Phys.
  Lett.\/} {\bf B738} 6--12 (\textit{Preprint} \eprint{1408.2517})

\bibitem{Pauk:2014rfa}
Pauk V and Vanderhaeghen M 2014 {\em Phys. Rev.\/} {\bf D90} 113012
  (\textit{Preprint} \eprint{1409.0819})

\bibitem{Masjuan:2012wy}
Masjuan P 2012 {\em Phys.Rev.\/} {\bf D86} 094021 (\textit{Preprint}
  \eprint{1206.2549})

\bibitem{Escribano:2013kba}
Escribano R, Masjuan P and Sanchez-Puertas P 2014 {\em Phys.Rev.\/} {\bf D89}
  034014 (\textit{Preprint} \eprint{1307.2061})

\bibitem{RamseyMusolf:2002cy}
Ramsey-Musolf M~J and Wise M~B 2002 {\em Phys. Rev. Lett.\/} {\bf 89} 041601
  (\textit{Preprint} \eprint{hep-ph/0201297})

\bibitem{Lepage:1980fj}
Lepage G~P and Brodsky S~J 1980 {\em Phys. Rev.\/} {\bf D22} 2157

\bibitem{Melnikov:2003xd}
Melnikov K and Vainshtein A 2004 {\em Phys. Rev.\/} {\bf D70} 113006
  (\textit{Preprint} \eprint{hep-ph/0312226})

\bibitem{Knecht:2001xc}
Knecht M and Nyffeler A 2001 {\em Eur. Phys. J.\/} {\bf C21} 659--678
  (\textit{Preprint} \eprint{hep-ph/0106034})

\bibitem{Benayoun:2015gxa}
Benayoun M, David P, DelBuono L and Jegerlehner F 2015  (\textit{Preprint}
  \eprint{1507.02943})

\bibitem{Roig:2014uja}
Roig P, Guevara A and L\'opez~Castro G 2014 {\em Phys. Rev.\/} {\bf D89} 073016
  (\textit{Preprint} \eprint{1401.4099})

\bibitem{Knecht:2001qf}
Knecht M and Nyffeler A 2002 {\em Phys. Rev.\/} {\bf D65} 073034
  (\textit{Preprint} \eprint{hep-ph/0111058})

\bibitem{Husek:2015wta}
Husek T and Leupold S 2015  (\textit{Preprint} \eprint{1507.00478})

\bibitem{'tHooft:1973jz}
't~Hooft G 1974 {\em Nucl. Phys.\/} {\bf B72} 461

\bibitem{Witten:1979kh}
Witten E 1979 {\em Nucl. Phys.\/} {\bf B160} 57

\bibitem{Masjuan:2007ay}
Masjuan P and Peris S 2007 {\em JHEP\/} {\bf 05} 040 (\textit{Preprint}
  \eprint{0704.1247})

\bibitem{Masjuan:2009wy}
Masjuan P and Peris S 2010 {\em Phys. Lett.\/} {\bf B686} 307--312
  (\textit{Preprint} \eprint{0903.0294})

\bibitem{Aubin:2012me}
Aubin C, Blum T, Golterman M and Peris S 2012 {\em Phys. Rev.\/} {\bf D86}
  054509 (\textit{Preprint} \eprint{1205.3695})

\bibitem{baker}
Baker G~A and Graves-Morris P 1996 {\em Pad{\'e} Approximants\/} 2nd ed ({\em
  Enciclopedia of Mathematics and its Applications\/} no~59) (New York:
  Cambridge University Press)

\bibitem{Escribano:2015nra}
Escribano R, Masjuan P and Sanchez-Puertas P 2015 {\em Eur. Phys. J.\/} {\bf
  C75} 414 (\textit{Preprint} \eprint{1504.07742})

\bibitem{Masjuan:2008fv}
Masjuan P, Peris S and Sanz-Cillero J~J 2008 {\em Phys. Rev.\/} {\bf D78}
  074028 (\textit{Preprint} \eprint{0807.4893})

\bibitem{Masjuan:2015lca}
Masjuan P and Sanchez-Puertas P 2015  (\textit{Preprint} \eprint{1504.07001})

\bibitem{Bijnens:2012hf}
Bijnens J, Kampf K and Lanz S 2012 {\em Nucl. Phys.\/} {\bf B860} 245--266
  (\textit{Preprint} \eprint{1201.2608})

\bibitem{Xiao:2015uva}
Xiao C~W, Dato T, Hanhart C, Kubis B, Mei{\ss}ner U~G and Wirzba A 2015
  (\textit{Preprint} \eprint{1509.02194})

\bibitem{Aguar-Bartolome:2013vpw}
Aguar-Bartolome P {\em et~al.\/} (A2) 2014 {\em Phys. Rev.\/} {\bf C89} 044608
  (\textit{Preprint} \eprint{1309.5648})

\bibitem{gm2}
Masjuan P and Sanchez-Puertas P  In preparation

\bibitem{Masjuan:2012qn}
Masjuan P and Vanderhaeghen M 2012  (\textit{Preprint} \eprint{1212.0357})

\bibitem{Drell:1959}
Drell S 1959 {\em Nuovo Cim.\/} {\bf 11} 693

\bibitem{Abouzaid:2006kk}
Abouzaid E {\em et~al.\/} (KTeV Collaboration) 2007 {\em Phys.Rev.\/} {\bf D75}
  012004 (\textit{Preprint} \eprint{hep-ex/0610072})

\bibitem{Dorokhov:2007bd}
Dorokhov A~E and Ivanov M~A 2007 {\em Phys. Rev.\/} {\bf D75} 114007
  (\textit{Preprint} \eprint{0704.3498})

\bibitem{Vasko:2011pi}
Vasko P and Novotny J 2011 {\em JHEP\/} {\bf 1110} 122 (\textit{Preprint}
  \eprint{1106.5956})

\bibitem{Husek:2014tna}
Husek T, Kampf K and Novotny J 2014 {\em Eur.Phys.J.\/} {\bf C74} 3010
  (\textit{Preprint} \eprint{1405.6927})

\bibitem{Kahn:2007ru}
Kahn Y, Schmitt M and Tait T~M 2008 {\em Phys.Rev.\/} {\bf D78} 115002
  (\textit{Preprint} \eprint{0712.0007})

\bibitem{Chang:2008np}
Chang Q and Yang Y~D 2009 {\em Phys.Lett.\/} {\bf B676} 88--93
  (\textit{Preprint} \eprint{0808.2933})

\bibitem{etaetapdecays}
Masjuan P and Sanchez-Puertas P  In preparation

\bibitem{Agakishiev:2013fwl}
Agakishiev G {\em et~al.\/} (HADES) 2014 {\em Phys.Lett.\/} {\bf B731} 265--271
  (\textit{Preprint} \eprint{1311.0216})

\bibitem{Abegg:1994wx}
Abegg R, Baldisseri A, Boudard A, Briscoe W, Fabbro B {\em et~al.\/} 1994 {\em
  Phys.Rev.\/} {\bf D50} 92--103

\bibitem{Achasov:2015mek}
Achasov M~N {\em et~al.\/} (SND) 2015 {\em Phys. Rev.\/} {\bf D91} 092010
  (\textit{Preprint} \eprint{1504.01245})

\bibitem{Akhmetshin:2014hxv}
Akhmetshin R {\em et~al.\/} (CMD-3) 2015 {\em Phys.Lett.\/} {\bf B740} 273--277
  (\textit{Preprint} \eprint{1409.1664})

\bibitem{Dorokhov:2009xs}
Dorokhov A, Ivanov M and Kovalenko S 2009 {\em Phys.Lett.\/} {\bf B677}
  145--149 (\textit{Preprint} \eprint{0903.4249})

\bibitem{Knecht:1999gb}
Knecht M, Peris S, Perrottet M and de~Rafael E 1999 {\em Phys.Rev.Lett.\/} {\bf
  83} 5230--5233 (\textit{Preprint} \eprint{hep-ph/9908283})

\end{thebibliography}
\end{document}